\documentclass[12pt]{article}\pdfoutput=1

\usepackage{amsmath}
\usepackage{amsfonts}
\usepackage{amssymb}
\usepackage{amsthm}
\usepackage{mathrsfs}
\usepackage{graphicx}
\usepackage{subfigure}
\usepackage{caption}             
\usepackage[usenames]{color}
\usepackage{verbatim}
\usepackage{appendix}
\usepackage[nottoc]{tocbibind}
\usepackage{sectsty}
\sectionfont{\normalsize}
\subsectionfont{\normalsize}
\renewcommand{\Im}{\operatorname{Im}}
\renewcommand{\Re}{\operatorname{Re}}
\usepackage{yfonts}
\newcommand{\wn}{\textswab{w}\kern1pt}
\newcommand{\qn}{\textswab{q}\kern1pt}
\usepackage{hyperref}

\title{\bf Shear channel correlators from hot charged black holes} 
\author{\normalsize Daniel K.~Brattan and Simon A.~Gentle\\ \\
        \small \it Centre for Particle Theory \& Department of Mathematical Sciences, \\
        \small \it Science Laboratories, South Road, Durham DH1 3LE, United Kingdom. \\ \\
        \normalsize\href{mailto:d.k.brattan@durham.ac.uk}{\texttt{d.k.brattan}}\texttt{, }\href{mailto:s.a.gentle@durham.ac.uk}{\texttt{s.a.gentle@durham.ac.uk}}}
\date{}

\begin{document}

\maketitle
\thispagestyle{empty}                        
\numberwithin{equation}{section}

\begin{abstract}
We compute numerically the full retarded Green's functions for conserved currents in the shear channel of a (2+1)-dimensional field theory at non-zero temperature and density.  This theory is assumed to be holographically dual to a non-extremal, electric Reissner-Nordstr\o m AdS$_4$ black hole with planar horizon.  Using the holographic description we obtain results for arbitrary frequencies and momenta and survey the detailed structure of these correlators.  In particular, we demonstrate the `repulsion' and `clover-leaf crossing' of their poles and stress the importance of the residues at the poles beyond the hydrodynamic regime.  As a consistency check, we show that our results agree precisely with existing literature for the appropriate quasinormal frequencies of the bulk theory.  
\end{abstract}

\pagebreak
\setcounter{page}{1}

\tableofcontents

\section{Introduction}\label{sec:Introduction}
The AdS/CFT correspondence is a valuable tool in the study of strongly-coupled systems.  For example, the real-time correlators of such systems can be extracted from classical gravity \cite{Son:2002sd,Herzog:2002pc}.  By tuning the physics of the bulk theory, the boundary theory can be made to mimic the qualitative features of many condensed matter theories.  Key examples are the superfluid phase transition \cite{Gubser:2008px,Hartnoll:2008vx,Hartnoll:2008kx} and a Fermi surface \cite{Liu:2009dm,Cubrovic:2009ye,Faulkner:2009wj}.  See \cite{Hartnoll:2009sz}, \cite{Horowitz:2010gk} and \cite{McGreevy:2009xe} for reviews.

Many previous studies have restricted to calculating just the locations of poles in retarded Green's functions.  These are found by computing the quasinormal frequencies of the bulk theory \cite{Birmingham:2001pj,Son:2002sd}.  As the characteristic oscillations of black objects, quasinormal modes typically decay exponentially in time (see \cite{Horowitz:1999jd} for an example and \cite{Berti:2009kk} for a review).  By searching for modes whose frequencies have instead a positive imaginary part, one can study the linear stability of the bulk theory and thus the likelihood of phase transitions in the boundary theory.  For example, the quasinormal frequencies for shear-type electromagnetic and gravitational perturbations of a Reissner-Nordstr\o m AdS$_4$ black hole were computed in \cite{Edalati:2010hk} and no such instabilites were found.  Many studies, even those in which the retarded Green's functions themselves are computed, often restrict to the regime of long wavelengths and/or small frequencies.  In this limit, the effective description is given by hydrodynamics and only the leading poles are significant \cite{Bhattacharyya:2008jc}.  Some examples are \cite{Policastro:2002se,Policastro:2002tn} and \cite{Kovtun:2005ev}.\footnote{See also \cite{Hubeny:2010ry} for a review of the recent developments in applying holographic methods to understand non-equilibrium physics in strongly-coupled field theories.}

In this paper we aim to go beyond these studies.  We compute the retarded Green's functions in full for conserved currents in the shear channel, considering arbitrary frequencies and momenta.  We consider a field theory which is assumed to be dual to an electric Reissner-Nordstr\o m AdS$_4$ black hole with planar horizon.  On general grounds, this field theory is expected to live in (2+1)-dimensional Minkowski space and have a $U(1)$ charge density.  We will work exclusively at non-zero temperature by studying a non-extremal black hole.  This calculation extends the work of \cite{Edalati:2010hk}.

We wish to explore the detailed structure of these correlators.  In this paper we show that their poles have varying residues and that this information can be used to assess the dominance of these poles beyond the hydrodynamic regime.   After analysing the properties of these correlators for large and small values of the boundary theory parameters, we examine the curious and intricate motion of their poles at intermediate parameter values.  

In the following section we outline the bulk theory and the prescription for computing the correlators of interest.  For more details we refer the reader to \cite{Edalati:2010hk}, whose conventions we share.  In Section~\ref{sec:Results} we present and discuss our results and the various consistency checks they satisfy.  We conclude with thoughts for future work.

\emph{Note added:} For comparison, we show in Appendix~\ref{app:Hydrodynamics} that we recover the hydrodynamics of the dual theory if we take the appropriate limit of our more general numerical results.   We also illustrate how the effective theory provided by hydrodynamics breaks down for shorter wavelengths and larger frequencies.  Our analytical results agree with those of \cite{Ge:2010yc}.

\section{Background and method}\label{sec:Background}
An action for a $U(1)$ gauge field $A$ coupled to gravity with a negative cosmological constant in four spacetime dimensions is
\begin{equation}\label{eq:action}
S=\frac{1}{2\kappa^2_4}\int d^4x \sqrt{-g} \left( R + \frac{6}{L^2} - L^2F_{ab}F^{ab} \right),
\end{equation}
where $L$ is the curvature radius and $F=dA$.  A solution to this theory describing an electric Reissner-Nordstr\o m AdS$_4$ black hole with planar horizon can be written
\begin{equation}\label{eq:background}
ds^2 = \frac{r^2}{L^2} \left( -fdt^2 + dx^2 +dy^2  \right)+ \frac{L^2}{r^2f}dr^2,\quad A = \mu\left(1 - \frac{r_0}{r}\right)dt,
\end{equation}
where
\begin{equation}\label{eq:fandmu}
f(r)=1-(1+Q^2)\left(\frac{r_0}{r}\right)^3 + Q^2\left(\frac{r_0}{r}\right)^4,\quad \mu=\frac{Qr_0}{L^2}.
\end{equation}
The (outer) black hole horizon is at $r=r_0$ and has temperature
\begin{equation}\label{eq:temperature}
T=\frac{\mu}{4\pi}\frac{3-Q^2}{Q}.
\end{equation}

Fluctuations about this background can be expressed in terms of gauge-invariant master fields $\Phi_{\pm}$ \cite{Kodama:2003kk}.  When written in terms of these fields, the linearised equations of motion decouple.  We are interested in the retarded Green's functions of operators in the shear channel of the boundary theory.  After using rotation invariance in the $(x,y)$ plane to set the momentum in the $y$ direction to zero, we can focus on correlators of $\hat{J}_y, \hat{T}_{xy}$ and $\hat{T}_{yt}$.  The master fields appropriate to this case, as well as further references and details, can be found in \cite{Edalati:2010hk}.

To aid numerical work, we introduce the dimensionless coordinate $z=r_0/r$.  In these coordinates the AdS$_4$ boundary is at $z=0$ and the black hole horizon is at $z=1$.  The equation for the master fields, given in \cite{Edalati:2010hk}, becomes
\begin{equation}\label{eq:masterfieldeom}
z^2f(f\Phi_{\pm}')'+\left[-zff'+Q^2z^2(\wn^2-\qn^2f)-2Q^2g_{\pm}z^3f\right]\Phi_{\pm}=0,
\end{equation}
where a prime denotes differentiation with respect to $z$ and
\begin{equation}\label{eq:g}
g_{\pm}(\qn)=\frac{3}{4}\left(1+\frac{1}{Q^2}\right)\left(1\pm\sqrt{1+\frac{16}{9}\left(1+\frac{1}{Q^2}\right)^{-2}\qn^2}\right).
\end{equation}
The dimensionless parameters $\wn$ and $\qn$ are respectively the frequency and spatial momentum of the perturbation in the boundary theory, as measured with respect to the chemical potential.  The latter is identified from the bulk theory to be $\mu$, as given in (\ref{eq:fandmu}).

To obtain the retarded Green's functions of interest, we must adapt the original prescription of \cite{Son:2002sd} to the case of coupled perturbations, considered in \cite{Son:2006em,Saremi:2006ep,Maeda:2006by}.  We must first solve (\ref{eq:masterfieldeom}) numerically, subject to infalling boundary conditions at the horizon.  Then we use the asymptotic fall-off of $\Phi_{\pm}(z)$ at the AdS$_4$ boundary, given by
\begin{equation}\label{eq:falloff}
\Phi_{\pm} \overset{z\to 0}{\sim} \hat{\Phi}_{\pm}\left(1 + \hat{\Pi}_{\pm}z + \ldots\right),
\end{equation}
to extract $\hat{\Pi}_{\pm}(\wn,\qn)$ from this solution.\footnote{More details and a numerical issue can be found in Appendix~\ref{app:Numerical}.}  The matrix of retarded Green's functions in the shear channel can be constructed from these functions.  Whilst the method can be found in \cite{Edalati:2010hk}, we present the results generalised to the non-extremal case here:
\begin{align}
G_{yt,yt} &= \frac{2 \qn^2 Q^2  (g_- \hat{\Pi}_+ - g_+ \hat{\Pi}_-)}{3 (g_+ - g_-)} \label{eq:firstcorrelator} \\
G_{xy,xy} &= \frac{2 \wn^2 Q^2  (g_- \hat{\Pi}_+ - g_+ \hat{\Pi}_-)}{3 (g_+ - g_-)} \label{eq:Gxyxy} \\
G_{xy,yt} &= -\frac{\qn \wn Q^2  (g_- \hat{\Pi}_+ - g_+ \hat{\Pi}_-)}{3 (g_+ - g_-)} \\
G_{yt,xy} &= -\frac{ \qn \wn Q^2  (g_- \hat{\Pi}_+ - g_+ \hat{\Pi}_-)}{3 (g_+ - g_-)} \\
G_{xy,y} &= \frac{2 \qn \wn Q^2   (\hat{\Pi}_+ - \hat{\Pi}_-)}{3 \mu (g_+ - g_-)} \label{eq:Gxyy} \\
G_{y,xy} &= \frac{2 \qn \wn   (\hat{\Pi}_+ - \hat{\Pi}_-)}{\mu (g_+ - g_-)}
\end{align}
\begin{align}
G_{yt,y} &= -\frac{2 \qn^2 Q^2   (\hat{\Pi}_+ - \hat{\Pi}_-)}{3 \mu (g_+ - g_-)} \\
G_{y,yt} &= -\frac{2 \qn^2  (\hat{\Pi}_+ - \hat{\Pi}_-)}{\mu (g_+ - g_-)} \\
G_{y,y} &= -\frac{8   (g_+\hat{\Pi}_+ - g_-\hat{\Pi}_-)}{\mu^2 (g_+ - g_-)}. \label{eq:lastcorrelator}
\end{align}
Note that a contact term has been suppressed in the $G_{yt,y}$ correlator and that all correlators are functions of $\wn$ and $\qn$.  Note also that we can pick the normalisation of two-point functions because we have the freedom to choose an operator basis on the boundary.  As such, we have chosen to normalise all correlators with respect to $G_{y,y}$.  These results reduce to those given in \cite{Edalati:2010hk} for zero temperature, i.e.\ for $Q^2=3$.

We set $L=1$ in our numerics and absorb a factor of $r_0$ into $(t,x,y)$.  As a result, $\mu=Q$.  In the rest of this paper, we refer to $\hat{\Pi}_{\pm}$ as `the retarded Green's functions for $\Phi_{\pm}$' for brevity.

\section{Results and discussion}\label{sec:Results}
\begin{figure}[htb]
\begin{center}
\includegraphics[width=0.9\textwidth]{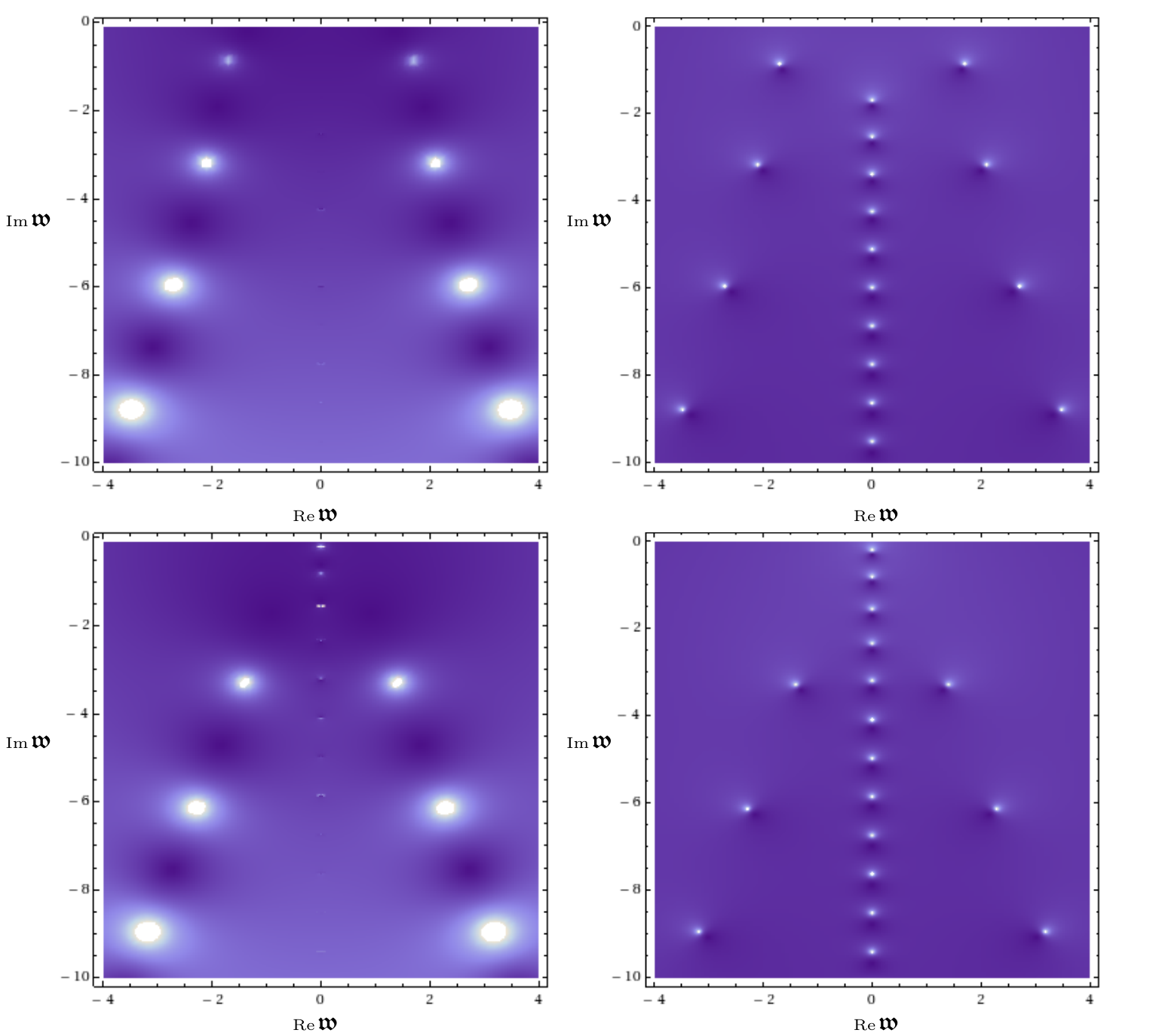}
\caption{\it A comparison between density plots of $|\hat{\Pi}_{\pm}|$ on the complex $\wn$ plane (left panels) and the quasinormal frequencies for $\Phi_{\pm}$.  The top row is for $\Phi_+$ and the bottom row is for $\Phi_-$.  All plots have $\qn=1$ and $T/\mu=0.09$.  As we discuss later, the on-axis modes are weaker but have been tested thoroughly against the quasinormal spectrum in a finer plot.}
\label{fig:RNAdS_Master}
\end{center}
\vskip -1em
\end{figure}

To ensure that the retarded Green's functions for $\Phi_{\pm}$ have been calculated correctly, we must check that the locations of their poles match the quasinormal spectrum for the appropriate bulk fluctuations \cite{Birmingham:2001pj,Son:2002sd}.  In Figure~\ref{fig:RNAdS_Master} we demonstrate that this is indeed the case for our results.  The quasinormal frequencies were computed using the determinant method pioneered in \cite{Leaver:1990zz} and explained in detail in \cite{Denef:2009yy}.\footnote{SAG would like to thank Sean Hartnoll for useful tips on generating these quasinormal spectra.}  We find precise agreement with the quasinormal frequency plots for non-zero temperature shown in \cite{Edalati:2010hk}.

First, a note on our terminology.  From the gravity perspective, modes with quasinormal frequencies on the negative $\Im\wn$ axis are associated with purely decaying perturbations of the black hole.  We refer to these as `on-axis' modes.  The `off-axis' modes have an additional oscillatory component.   

\begin{figure}[htb]
\begin{center}
\includegraphics[width=0.9\textwidth]{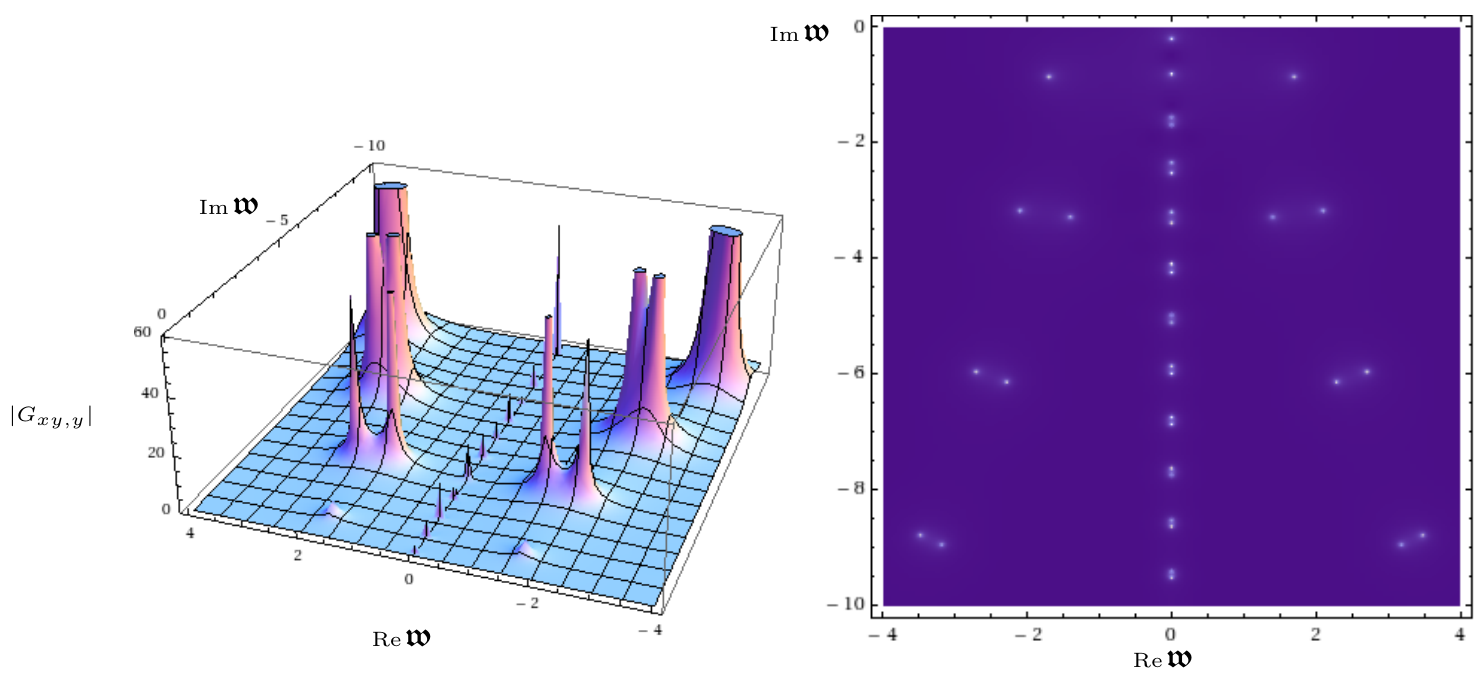}
\caption{\it A comparison between a surface plot of $|G_{xy,y}|$ on the complex $\wn$ plane (left) and the appropriate quasinormal frequencies.  Both plots have $\qn=1$ and $T/\mu=0.09$.}
\label{fig:Gxyy}
\end{center}
\vskip -2em
\end{figure}

As an example, we present in Figure~\ref{fig:Gxyy} the full result for the $G_{xy,y}$ correlator, given in (\ref{eq:Gxyy}).  It can be seen from (\ref{eq:firstcorrelator}-\ref{eq:lastcorrelator}) and (\ref{eq:g}) that the poles in the various correlators are determined completely by $\hat{\Pi}_{\pm}$, which we now study for various parameter ranges. 

\subsection{Residues}\label{ssec:Residues}
\begin{figure}[htb]
\begin{center}
\includegraphics[width=0.96\textwidth]{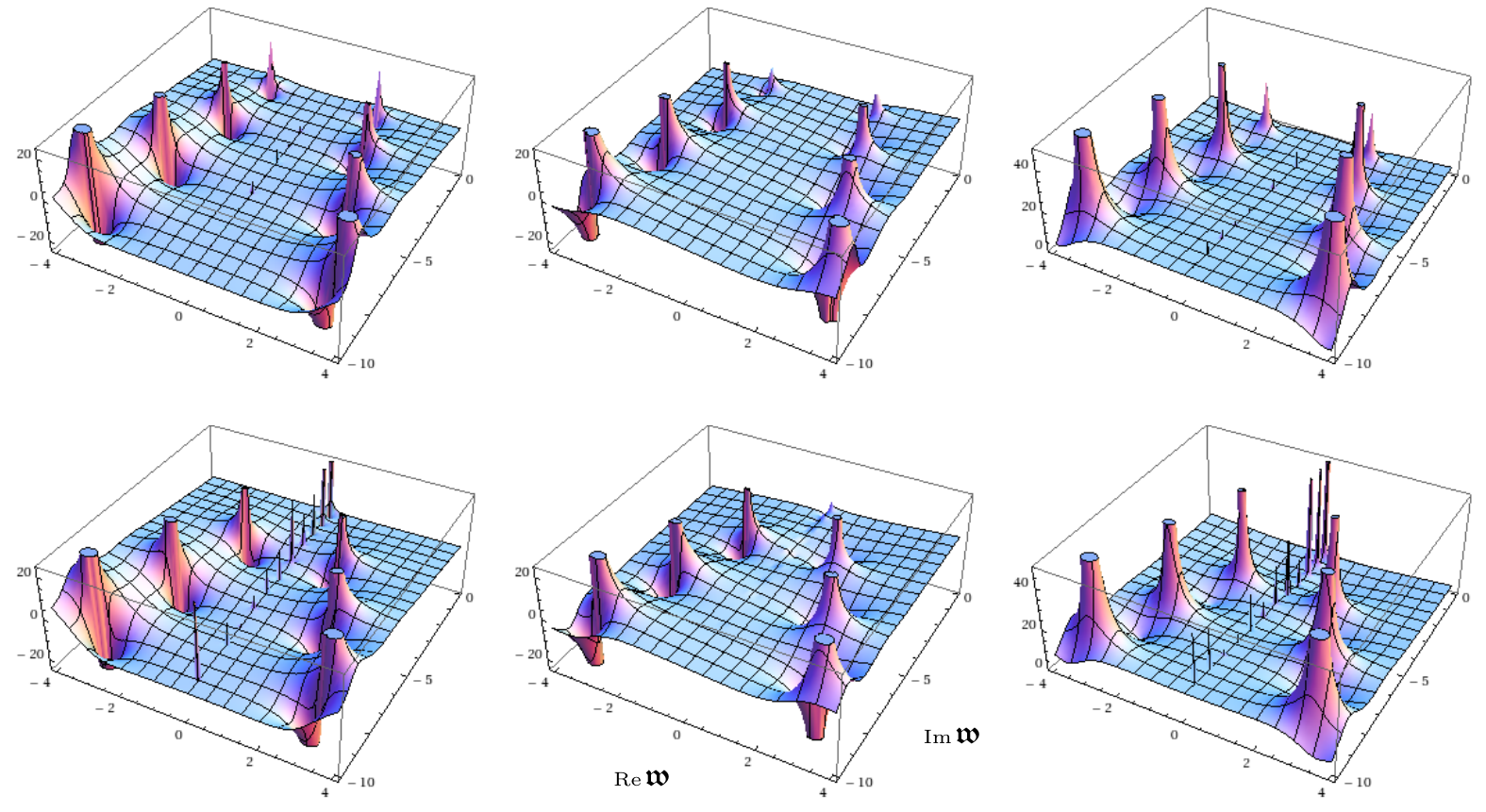}
\caption{\it Surface plots of $\hat{\Pi}_{+}$ (top row) and $\hat{\Pi}_{-}$ (bottom row) on the complex $\wn$ plane at $\qn=1$, $T/\mu=0.09$.  We show (from left to right) the real part, imaginary part and absolute value of each.  All plots have the same orientation on the plane, as indicated.}
\label{fig:PiPlusMinusSurfacePlots}
\end{center}
\vskip -2em
\end{figure}

In Figure~\ref{fig:PiPlusMinusSurfacePlots} we show surface plots of $\hat{\Pi}_{\pm}$ on the complex $\wn$ plane.  They verify the usual symmetries of retarded Green's functions, namely
\begin{equation}\label{eq:symmetry}
\overline{\hat{\Pi}_{\pm}(\wn,\qn)} = \hat{\Pi}_{\pm}(-\bar{\wn},\qn),
\end{equation}
which can be shown to follow from the $\wn \leftrightarrow - \bar{\wn}$ symmetry of the equation of motion (\ref{eq:masterfieldeom}) and the infalling boundary conditions.  

Note that the off-axis poles have much wider peaks than those on-axis, indicating larger residues.  We would like to understand how important these residues are in determining the behaviour of the boundary theory.  Suppose we assume that $\hat{\Pi}_{\pm}$ have only simple poles which are well-separated (an assumption we question in a later subsection).  Then they will have the form
\begin{equation}\label{eq:GRassumoverpoles}
\hat{\Pi}_{\pm}(\wn,\qn) = \sum_{i} \frac{R_{i}\left(\qn;T/\mu \right)}{\wn - \wn_{i}\left(\qn;T/\mu \right)} + \mathrm{ analytic \; pieces},
\end{equation}
where the locations of the poles $\wn_{i}$ in the complex $\wn$ plane have real part $\tilde{\wn}_{i}(\qn;T/\mu)$ and imaginary part $ -  \Gamma_{i}(\qn;T/\mu)$.  By performing a Fourier transform, we show in Appendix~\ref{app:Retarded} that the quantity
\[
R_{i}(\qn;T/\mu) e^{-\mu\Gamma_i(\qn;T/\mu)(t-t')}
\]
determines the contribution of a given pole to $\hat{\Pi}_{\pm}$.  Beyond the hydrodynamic regime, i.e.\ for $t-t' \lesssim 1$, the residues can therefore be cruicial in determining which poles dominate the linear response.

\subsection{Behaviour at large frequency and momentum} \label{ssec:Behaviour}
\begin{figure}[htb]
\begin{center}
\includegraphics[width=0.84\textwidth]{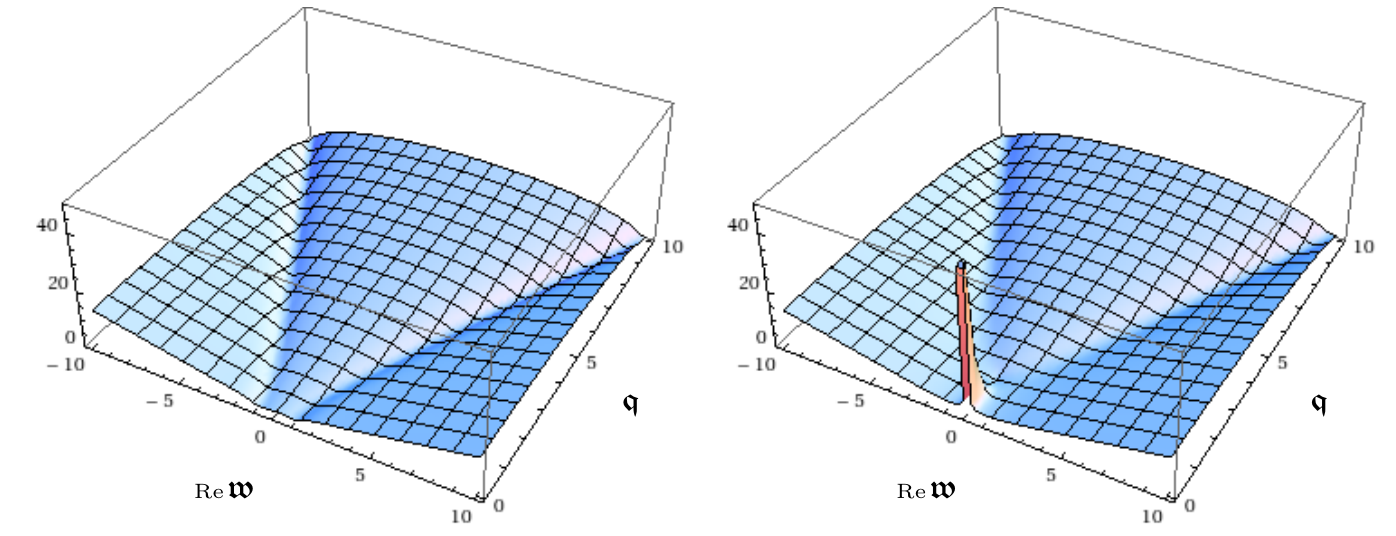}
\caption{\it Surface plots of $| \hat{\Pi}_{+} |$ (left) and $| \hat{\Pi}_{-} |$ (right) on the $(\Re\wn,\qn)$ plane at $T/\mu=0.09$ to demonstrate the large $\wn$ and $\qn$ scaling.  Note that we avoid $\qn=0$: in that case the symmetry is enhanced and so the master fields are different, as discussed in \normalfont\cite{Edalati:2010hk}.}
\label{fig:AbsPiOmegaKSurface}
\end{center}
\vskip -2em
\end{figure}

At large $\wn$ and $\qn$ we can perform a WKB analysis of the equation of motion.  This yields
\begin{equation}\label{eq:PiPlusMinusLarge}
\hat{\Pi} = \pm i Q \sqrt{\wn^2 - \qn^2}
\end{equation}
for both $\hat{\Pi}_{\pm}$, where we consider real $\wn$.  So for $\wn \gg \qn$ or $\qn \gg \wn$ the growth of $|\hat{\Pi}_{\pm}|$ is linear in the larger parameter, which we demonstrate in Figure~\ref{fig:AbsPiOmegaKSurface}.  We note the zeroes at $\wn=\pm\qn$ for large $\Re\wn$ and $\qn$ in confirmation of our result and also the strong presence of the hydrodyanmic pole in $ \hat{\Pi}_{-} $, to be discussed in a Section~\ref{ssec:Pole}.

We have found the full retarded Green's functions for a non-trivial density matrix in the boundary theory.  To isolate the modification from the vacuum, i.e.\ the effect of adding the black hole to the bulk theory, one can subtract off this asymptotic behaviour.  However, this comes at the price of introducing the square root branch cut shown above.  The cut originates from the retarded Green's functions of the CFT dual to pure AdS$_4$, a fact which has been obscured somewhat by moving to master fields.

\subsection{Spectral function}
For a field theory in thermal equilibrium, the spectral function gives the number density of states in the ensemble with a particular $\wn$ and $\qn$.  The spectral function is defined up to normalisation by
    \begin{equation}
        \rho\left( \wn, \qn \right) = - \Im G_{R}\left(\wn,\qn \right)
    \end{equation}
for real $\wn$, where $G_R$ is the appropriate retarded Green's function.  

\begin{figure}[htb]
\begin{center}
\includegraphics[width=0.9\textwidth]{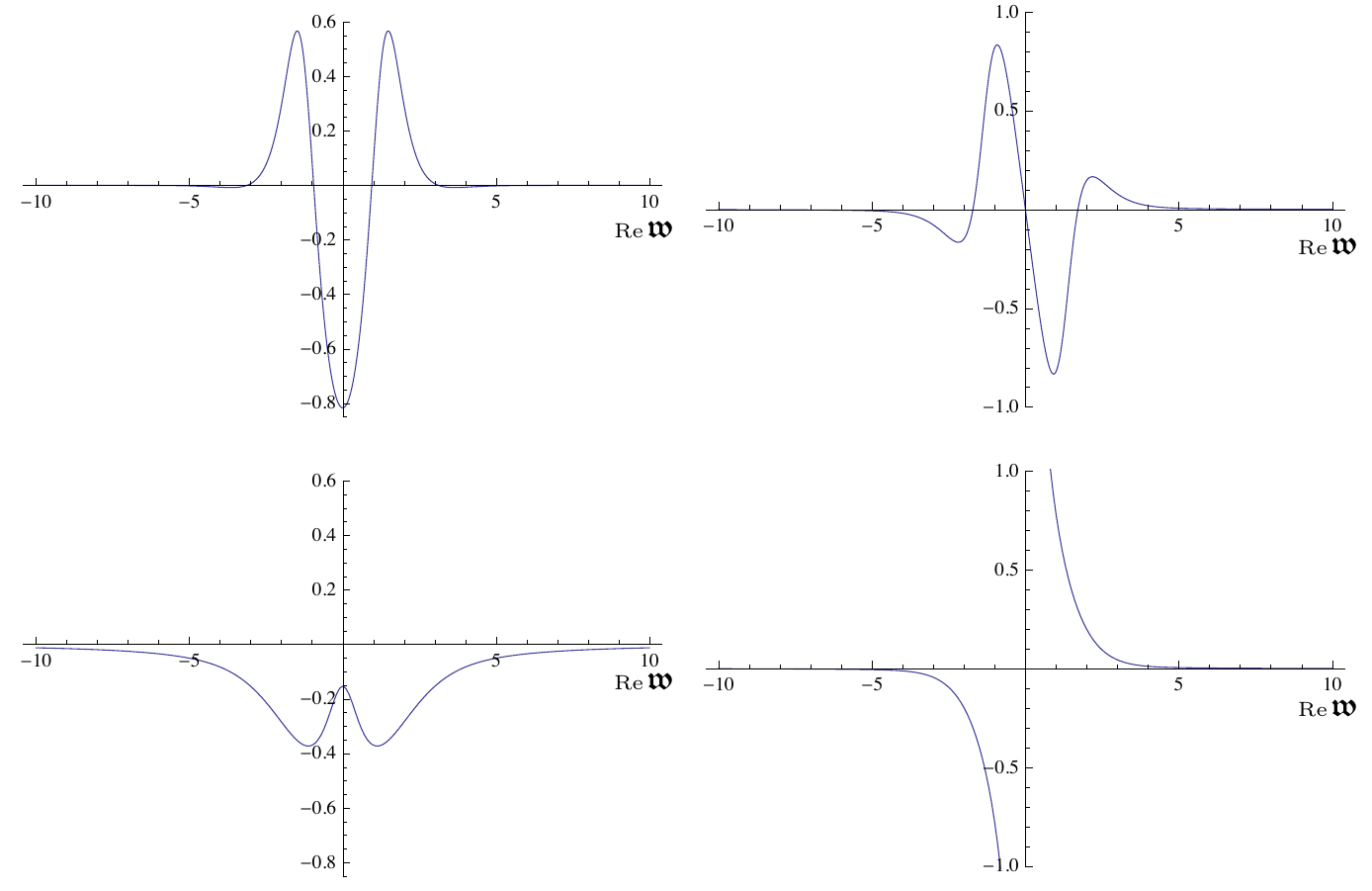}
\caption{\it Slices through the $(\Re\wn,\qn)$ plane of $\hat{\Pi}_{+}$ (top row) and $\hat{\Pi}_{-}$ (bottom row) at $\qn=10^{-6}$ and $T/\mu=0.09$.  The left and right panels show the real and imaginary parts, respectively.}
\label{fig:Omega_plots_at_fixed_k}
\end{center}
\vskip -2em
\end{figure}

The features on the $\Re\wn$ axis can be explained by studying the general form (\ref{eq:GRassumoverpoles}).  Making use of the symmetry (\ref{eq:symmetry}) we consider a straightforward re-writing of this expression, with dependence on $\qn$ and $T/\mu$ suppressed:
    \begin{equation}
        \hat{\Pi}_{\pm}(\wn) = \sum_{i} \left[ \frac{a_{i}+ib_{i}}{\wn - |\tilde{\wn}_i|+i\Gamma_{i}} -  \frac{a_{i}-ib_{i}}{\wn + |\tilde{\wn}_i|+i\Gamma_{i}} \right] + \sum_{j} \frac{ic_{j}}{\wn+i\Gamma_{j}} + \mathrm{ analytic}.
    \end{equation}
Here, $i$ runs over one half of the off-axis poles and $j$ runs over the on-axis poles.  Taking real and imaginary parts of the above form we obtain 
 \begin{align}
        \Re \hat{\Pi}_{\pm} &= \sum_{i} \left[\frac{a_{i}\wn-\left(a_{i}|\tilde{\wn}_i|-b_{i}\Gamma_{i}\right)}{\left(\wn - |\tilde{\wn}_i|\right)^2 + \Gamma_{i}^2}
                                                   - \frac{a_{i}\wn+\left(a_{i}|\tilde{\wn}_i|-b_{i}\Gamma_{i}\right)}{\left(\wn + |\tilde{\wn}_i|\right)^2 + \Gamma_{i}^2} \right]
                                                   + \sum_{j} \frac{c_{j}\Gamma_{j}}{\wn^{2} + \Gamma_{j}^2} + \mathrm{analytic}  \\
        \Im \hat{\Pi}_{\pm} &= \sum_{i} \left[\frac{b_{i}\wn-\left(a_{i}\Gamma_{i}+b_{i}|\tilde{\wn}_i|\right)}{\left(\wn-|\tilde{\wn}_i|\right)^2+\Gamma_{i}^2}
                                                   + \frac{b_{i}\wn+\left(a_{i}\Gamma_{i}+b_{i}|\tilde{\wn}_i|\right)}{\left(\wn+|\tilde{\wn}_i|\right)^2+\Gamma_{i}^2} \right]
                                                   + \sum_{j} \frac{ c_{j} \wn }{\wn^{2} + \Gamma_{j}^2} + \mathrm{analytic}.
    \end{align}
In Figure~\ref{fig:Omega_plots_at_fixed_k} we have taken a slice of the $(\Re\wn,\qn)$ plane at $\qn=10^{-6}$.  The expressions above take their largest values whenever $\wn=0$ or $\pm |\tilde{\wn}_i|$, which is indeed what we find in Figure~\ref{fig:Omega_plots_at_fixed_k}.  In effect, the presence of poles lower down in the complex $\wn$ plane is `projected' onto the retarded Green's functions at real $\wn$.  

Note that in Figure~\ref{fig:Omega_plots_at_fixed_k} we have subtracted the large $\wn$ and $\qn$ behaviour discussed previously in order to reveal the peaks that occur away from $\wn=0$.  Our numerics confirm that the modified retarded Green's functions tend to zero as $\Re\wn$ increases.  We have chosen to subtract off the positive branch.

By considering $\qn\ll 1$ we are working in the long-wavelength regime.  If instead we consider $\qn\gg1$, for which the spatial perturbations are much smaller than the scales in the boundary theory, the poles on the $\Im\wn$ axis move much further down the complex $\wn$ plane and their effect on the spectral function diminishes.

\subsection{Pole motion}\label{ssec:Pole}
As noted in \cite{Edalati:2010hk}, the lowest quasinormal frequency for $\Phi_-$ approaches zero as $\qn\to 0$.  We found the same behaviour for the lowest pole in $\hat{\Pi}_-$; as such, this can be identified as the hydrodynamic pole.  The behaviour of the poles as $\qn$ and $T/\mu$ are varied is particularly rich and, to our knowledge, has not been studied in detail before.

\begin{figure}[htb]
\begin{center}
\includegraphics[width=0.95\textwidth]{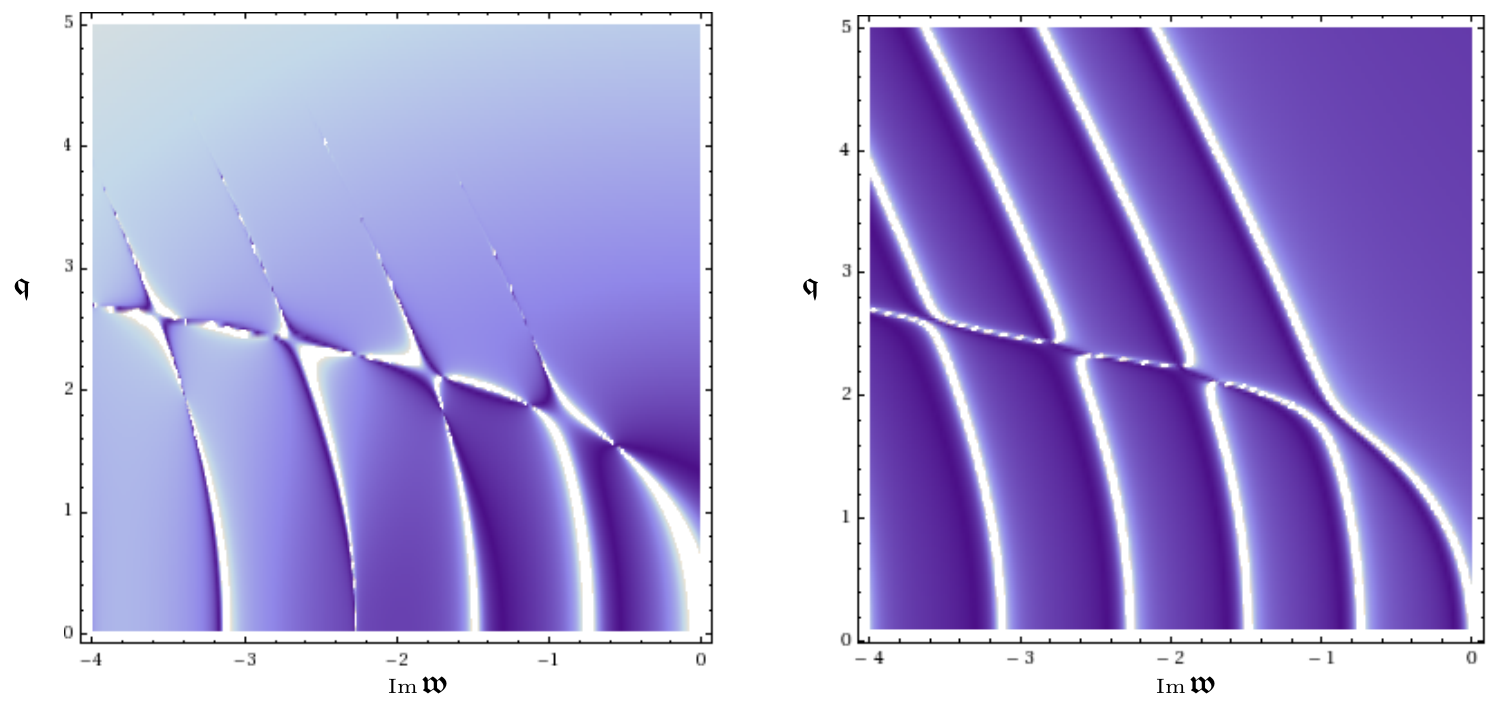}
\caption{\it Trajectories on the $(\Im\wn,\qn)$ plane of the on-axis poles of $\hat{\Pi}_{-}$ (left) and the motion of the associated quasinormal frequencies.  We fix $T/\mu=0.09$ and consider $0.1\leq\qn\leq 5$.}
\label{fig:AxisPoleTrajectories}
\end{center}
\vskip -2em
\end{figure}

The trajectories of the on-axis poles of $\hat{\Pi}_{-}$ at fixed $T/\mu$ are shown in Figure~\ref{fig:AxisPoleTrajectories}.  As $\qn$ is increased, all these poles move down the $\Im\wn$ axis.  The on-axis poles of $\hat{\Pi}_+$ have very simple trajectories and so we focus on the behaviour of $\hat{\Pi}_-$.   

Firstly we note the apparent continuity of the white line characterising the hydrodynamic pole.  The physical implication of this behaviour is that in the hydrodynamic limit, where this is the only relevant pole, the dispersion relation of the quasi-particle corresponding to this pole is a smooth function of $\qn$.  This is given by  
\begin{equation}
\wn  = - i \mu D \qn^2  
\end{equation}
with diffusion constant $D$.\footnote{See Appendix~\ref{app:Hydrodynamics} for a discussion of the diffusion constant.}  In the left of Figure~\ref{fig:AxisPoleTrajectories} we can see there are values of $\qn$ for which a given on-axis mode disappears, as indicated by a gap in the associated trajectory (for example at $\qn \approx 1.6$ and $\wn \approx -0.6 i$).  If (\ref{eq:GRassumoverpoles}) holds, then we must conclude that the residue vanishes at these $(\wn,\qn)$.  With this in mind we see that, for finely tuned $\qn$, it may be possible for a previously sub-leading pole to give the dominant contribution to $\hat{\Pi}_{-}$.  Note that whilst the quasinormal plots can show this continuity, full knowledge of the correlators is required to determine which pole dominates.

Also, we observe that the trajectories of all on-axis poles become less pronounced as $\qn$ is increased, which indicates that the residues of the poles become smaller.   Furthermore, there are no poles at small $|\wn|$ for sufficiently large $\qn$.  

The `interactions' between the poles of $\hat{\Pi}_{-}$ are quite intricate as $T/\mu$ and $\qn$ are varied.  These features are best displayed in animations, but we give a brief summary here.\footnote{Animations are available at \href{http://www.maths.dur.ac.uk/~rcqn58/research.html}{\texttt{http://www.maths.dur.ac.uk/$\sim$rcqn58/research.html}}.  These show the motion of the quasinormal frequencies for $\Phi_-$.  We observe the same behaviour for $\hat{\Pi}_{-}$ and as such it is not an artifact of using a matrix of finite size in the determinant method.  These animations also show the behaviour of the off-axis poles, which remains to be understood.}

We observe two types of behaviour: `repulsion' and and `clover-leaf crossing'.  Suppose we fix $T/\mu$ and increase $\qn$.  All poles on the $\Im\wn$ axis move downwards but the hydrodynamic pole moves down further for the same change in $\qn$, as shown in Figure~\ref{fig:AxisPoleTrajectories}.  It slows down as it approaches the next pole and the latter speeds up but the two never touch, just as for two like magnetic poles being brought together.  As this second pole approaches the third, the two coalesce into a single object, split and move off-axis, coalesce a second time then split and move on-axis in a clover-leaf pattern.  

\begin{figure}[htb]
\begin{center}
\includegraphics[width=0.5\textwidth]{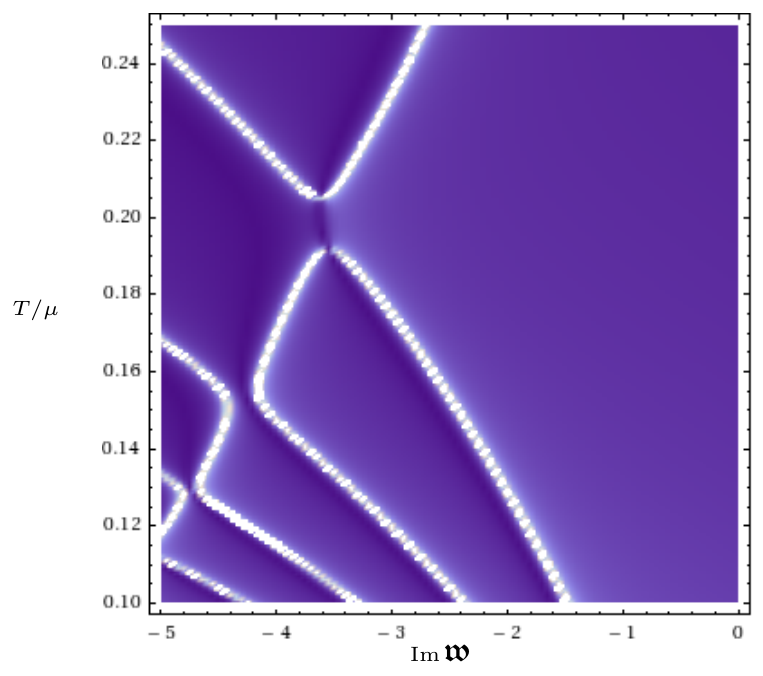}
\caption{\it Trajectories on the $(\Im\wn,T/\mu)$ plane of the on-axis quasinormal frequencies of $\Phi_-$.  We fix $\qn=3$ and consider $0.1\leq T/\mu\leq 0.25$.}
\label{fig:minus_trajectories_varyT}
\end{center}
\vskip -2em
\end{figure}

\begin{figure}[htbp]
\begin{center}
\includegraphics[width=0.95\textwidth]{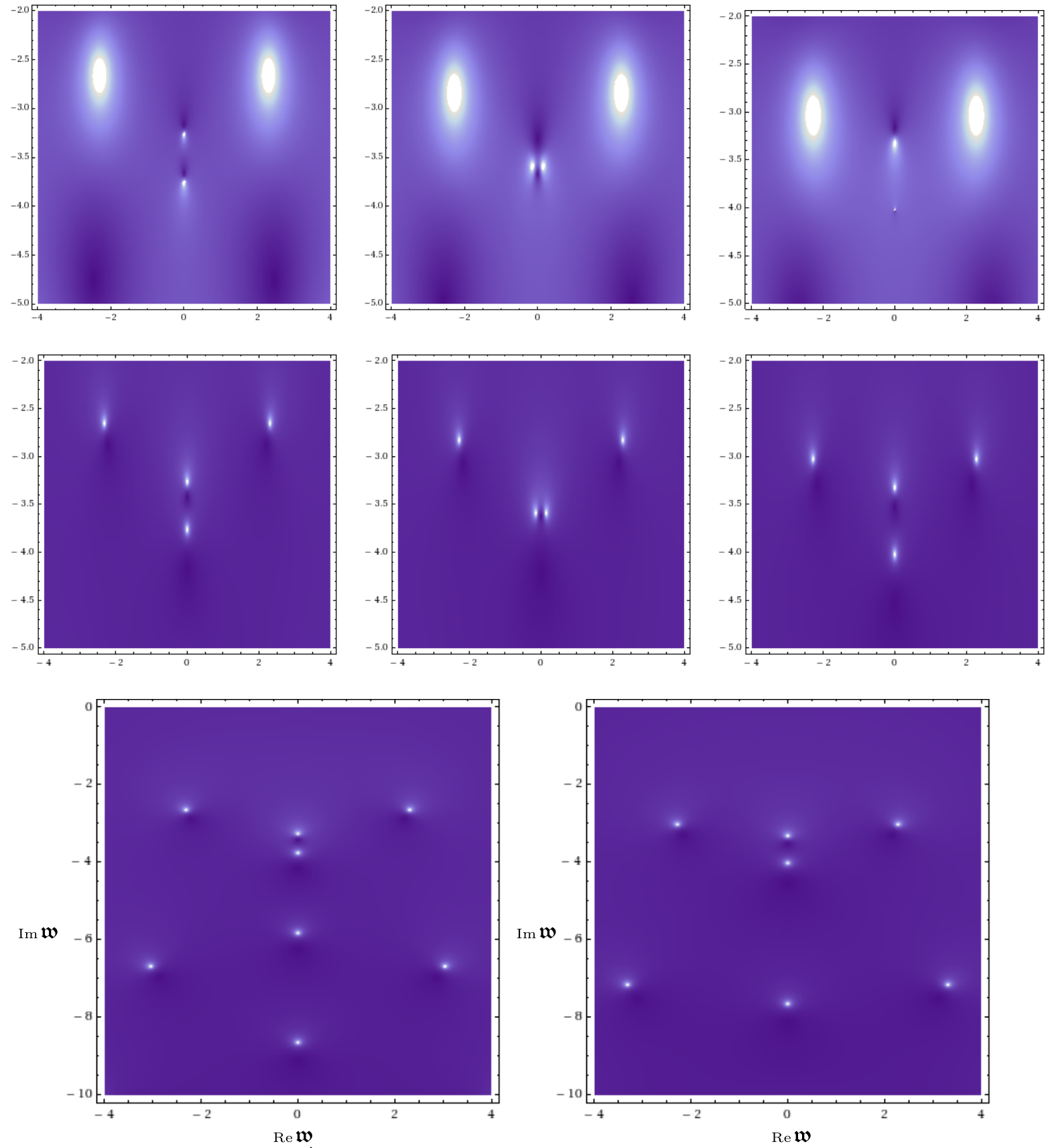}
\caption{\it Density plots of $|\hat{\Pi}_{-}|$ on the complex $\wn$ plane at fixed $\qn=3$ are shown in the upper row.  The values of $T/\mu$ are (from left to right) $0.185$, $0.2$ and $0.215$.  These plots have been cropped to include only the poles which move off-axis and the leading poles.  For comparison, the quasinormal frequencies at the same values of parameters are given in the second row.  The final row contains two quasinormal frequency plots, one at $T/\mu=0.185$ and one at $T/\mu=0.215$.  These show a larger region of the complex $\wn$ plane, including the origin, for orientation.}
\label{fig:CrossingandClover}
\end{center}
\end{figure}

Now suppose we fix $\qn$ at some larger value and increase $T/\mu$.  Again, all the poles move downwards as $T/\mu$ is increased, except there is always one pole which moves upwards.  An example of this is shown in Figure~\ref{fig:minus_trajectories_varyT} and a sample crossing is depicted in Figure~\ref{fig:CrossingandClover}.  We also see several repulsions, so it appears that a different pole moves upwards each time.

We have so far not been able to tag the poles in a clover-leaf crossing and track their motion independently.  The fact that the poles coalesce follows from the symmetry (\ref{eq:symmetry}): the poles must be symmetric about the $\Im\wn$ axis, so they cannot move round each other but must cross.  However, we do not know if this forms a sum of simple poles or a double pole.  If the latter occurs then (\ref{eq:GRassumoverpoles}) must be modified, so it would be interesting to map the behaviour of $\hat{\Pi}_-$, or even just the hydrodynamic pole, over a larger region of $(\qn,T/\mu)$ space. 

From the gravity perspective, we have shown that certain quasinormal modes switch from purely decaying to partially oscillatory behaviour as the parameters are varied.  For example, for the central plot of Figure~\ref{fig:CrossingandClover} there exist two modes degenerate in energy which move in opposite directions.  Now, those modes closest to the $\Re\wn$ axis dominate the late-time response of the black hole.  By studying this crossover we are exploring the intermediate response for perturbations of various $\qn$ of black holes with different $T/\mu$.  From the field theory perspective, the dispersion relation for the excitations associated with these poles develops a real part at this crossover.

\section{Outlook}\label{sec:Outlook}
In this paper we have considered a field theory which is assumed to be dual to a non-extremal Reissner-Nordstr\o m AdS$_4$ black hole.  We have computed numerically the full retarded Green's functions for conserved currents in the shear channel of this theory for non-trivial density matrix.  Our results are consistent with those of \cite{Edalati:2010hk}, but we have gone beyond that treatment to show many interesting features.  In particular, the linear response of the boundary theory exhibits rather peculiar behaviour beyond the hydrodynamic regime which we would like to understand in more detail.

An immediate application of our methods is to study the sound channel correlators of this non-zero temperature field theory.  This case was not considered in \cite{Edalati:2010pn}.  We would also like to compute the full retarded Green's functions for conserved currents in the shear channel at zero temperature.  We expect qualitatively different behaviour at zero temperature.  It has been argued in many examples (see \cite{Denef:2009yy}, for instance) that the double zero in $f(r)$ at the horizon of extremal Reissner-Nordstr\o m AdS$_4$ leads to a branch cut in the appropriate retarded Green's function(s) of the boundary theory.  This was first seen explicitly at small frequencies in \cite{Faulkner:2009wj}.  Whilst numerical results were reported in \cite{Edalati:2010pn} for the sound channel at zero temperature, only the small-$\wn$ behaviour was reported in \cite{Edalati:2010hk} for the shear channel, based on the analytical methods developed in \cite{Faulkner:2009wj}.  We would like to go beyond this regime, for which numerical methods are required.  However, the double zero in $f(r)$ shows up as an irregular singular point in the master field equations, leading to difficulties in the numerical computation of the retarded Green's functions by our methods.  It would be interesting to adapt our methods to tackle zero temperature for the shear channel.

Note that we are taking a `bottom-up' approach to the AdS/CFT correspondence.  As such, the boundary theory is not well-defined and we may miss some interesting features, particularly at low temperatures.  The hope is that such toy bulk theories can be embedded in a consistent truncation of ten- or eleven-dimensional supergravity.  Important milestones in this regard are \cite{Gauntlett:2009dn,Gauntlett:2009bh,Gubser:2009qm}.  We believe our techniques will be useful in exploring such setups.

\section*{Acknowledgements}
We would like to thank our advisor Mukund Rangamani for inspiration and guidance throughout this work.  It is also a pleasure to thank Pau Figueras, Sean Hartnoll, Patrick Kerner and Julian Sonner for very useful discussions and/or correspondence.  We are both supported by STFC studentships.

\appendix

\section{Retarded Green's functions in position space}\label{app:Retarded}
In this Appendix we derive the position space representation of the general form for $\hat{\Pi}_{\pm}$ given in (\ref{eq:GRassumoverpoles}).  First, consider the simpler function
\begin{equation}
\hat{\Pi}_{\pm}(\wn,\qn) = \frac{R(\qn)}{\wn - \tilde{\wn}(\qn) + i \Gamma(\qn)},
\end{equation}
where we have suppressed the dependence on $T/\mu$ for clarity.  Fourier transforming with respect to $\omega$ and $k$ we find
\begin{align}
\hat{\Pi}_{\pm}(x-x') &= \mu \int \frac{d\omega dk}{\left(2\pi\right)^{2}}  \frac{R(\qn)}{\omega - \tilde{\omega}(\qn) + i \mu \Gamma(\qn)} e^{-i \omega (t-t') + i \vec{k} \cdot (\vec{x}-\vec{x}')} \nonumber \\
      &= \mu \int \frac{da dk}{\left(2\pi\right)^{2}} \frac{R(\qn)}{a} e^{-i \left(a+\tilde{\omega}(\qn) - i \mu \Gamma(\qn) \right) (t-t') + i \vec{k} \cdot (\vec{x}-\vec{x}')} \nonumber \\
      &= \mu \int \frac{dk}{\left(2\pi\right)} R(\qn) \left[ \frac{da}{2\pi} \frac{1}{a} e^{-i a(t-t')} \right] e^{-i \left( \tilde{\omega}(\qn) - i \mu \Gamma(\qn) \right) (t-t') + i \vec{k} \cdot (\vec{x}-\vec{x}')}. \nonumber
\end{align}
We can integrate the factor in square brackets using contour integration.  First we nudge the pole a short distance into the lower half-plane.  The contribution along the curved section of a semi-circular contour will vanish for different ranges of $t-t'$.  We recover the following result for the above integral over $a$:
\begin{align}
  \lim_{\epsilon \to 0} \int \frac{da}{2\pi} \frac{1}{a+i\epsilon} e^{-i a(t-t')} &=  \lim_{\epsilon \to 0}  \left\{ \begin{array}{cl} - 2 \pi i \lim_{a \to -i\epsilon} \frac{1}{2\pi} e^{-i a(t-t')},
              & (t-t')>0 \\
              0,
              & (t-t')<0
            \end{array} \right. \nonumber \\
    &= \lim_{\epsilon \to 0}  \left\{ \begin{array}{cl}
              - i e^{-(t-t')\epsilon},
              & (t-t')>0 \\
              0,
              & (t-t')<0
            \end{array} \right. \nonumber \\
    &= - i \theta(t-t'). \nonumber
\end{align}
Thus,
\begin{equation}
\hat{\Pi}_{\pm}(x-x') = - i \mu \theta(t-t') \int \frac{dk}{\left(2\pi\right)} R(\qn) e^{- \mu \Gamma(\qn)(t-t')} e^{-i \tilde{\omega}(\qn) (t-t') + i \vec{k} \cdot (\vec{x}-\vec{x}')}.
\end{equation}

The retarded Green's function of (\ref{eq:GRassumoverpoles}) is a sum over poles.  We apply the above procedure to each term in that sum to obtain
\begin{equation}
\hat{\Pi}_{\pm}(x-x') = - i \mu \theta(t-t') \int \frac{dk}{\left(2\pi\right)} \sum_i \left[R_i(\qn) e^{- \mu \Gamma_i(\qn)(t-t')} e^{-i \tilde{\omega}_i(\qn) (t-t')}\right]e^{  i \vec{k} \cdot (\vec{x}-\vec{x}')}.
\end{equation}
It is clear from this result that the contribution of each pole to $\hat{\Pi}_{\pm}$ is characterised by the quantity $R_i(\qn) e^{- \mu \Gamma_i(\qn)(t-t')}$, as claimed in Section~\ref{ssec:Residues}.

\section{Numerical method}\label{app:Numerical}

In this Appendix we give details of the method used to extract the retarded Green's functions for $\Phi_{\pm}$ numerically.  As outlined in Section~\ref{sec:Background}, we must first solve (\ref{eq:masterfieldeom}).  This is a second-order, linear ODE with a regular singular point at $z=1$ (the horizon).   We choose the following ansatz, where $\Phi$ denotes one of the $\Phi_{\pm}$:
\begin{equation}\label{eq:ansatz}
\Phi(z) = (z-1)^{-i \wn \mu /4\pi T} \phi(z).
\end{equation}
The first factor imposes the infalling boundary condition at $z=1$.  A unique solution is then specified by $\phi(1)$, which we are free to choose because the equation is linear.  We expand $\phi$ about $z=1$ up to some order, $N$, to generate the initial condition for a Runge-Kutta algorithm at some $z=1-\epsilon$.  To extract $\hat{\Pi}_{\pm}(\wn,\qn)$, we integrate out to the boundary and match this numerical solution to the asymptotic expansion given in (\ref{eq:falloff}).  The matching can be  performed using a root-finding method, for example.  

The main numerical issue comes from the initial condition.  Naively, we would like to choose a small $\epsilon$ so that the series expansion of $\phi$ is accurate with only a few terms.  However, we found that the pattern of  poles was completely washed out below a certain line in the complex $\wn$ plane.  This numerical instability appears if $\epsilon$ is small for $\Im\wn$ large and negative (and/or for $T/\mu$ small).   To see why, note that the ansatz (\ref{eq:ansatz}) for $\Phi$ at $z=1-\epsilon$ contains the factor
\[
\epsilon^{-i \wn \mu /4\pi T}.
\]
This factor becomes very large in these regimes if $\epsilon$ is small, leading to round-off errors.  

Thus, a larger $\epsilon$ must be chosen in these regimes.  As a consequence, a sufficiently large $N$ must be chosen to offset the error from starting the integration further from $z=1$.\footnote{SAG would like to thank Patrick Kerner for discussions on this issue.  See \cite{Kaminski:2009ce} for further evidence.}  The values of $\epsilon, N$ can  be constrained by ensuring that the locations of the poles match the quasinormal spectrum for the appropriate bulk fluctuations, as stated in Section~\ref{sec:Results}.

\section{Hydrodynamics}\label{app:Hydrodynamics}
In this Appendix we show that we recover the correct hydrodynamics if we take the appropriate limit of our numerical results.  We begin by reproducing the analytical expressions for $\hat{\Pi}_{\pm}$ found in \cite{Ge:2010yc} for the benefit of the reader.

As discussed in Section~\ref{sec:Background}, the matrix of corrrelators can be constructed from $\hat{\Pi}_{\pm}$.  We now compute the latter to leading order in $\wn$ and $\qn^2$ in the spirit of \cite{Policastro:2002se}.\footnote{There can be no odd powers of $\qn$ due to parity.}  Firstly, as in \cite{Ge:2010yc}, we consider the following field redefinition, where $\Phi$ denotes one of the $\Phi_{\pm}$:
\begin{equation}
\Phi(z) = (z-1)^{-i \wn \mu /4\pi T} F(z) G(z).
\end{equation}
The aim is to choose $G$ such that $F$ tends to an $\wn$- and $\qn$-independent constant, $F_0$, in the hydrodynamic limit $\wn,\qn\to 0$.  This choice greatly simplifies the subsequent analysis, as we will soon show, and leads to the equation
\begin{equation}
zfG''+zf'G'-(f'+2Q^2g_{\pm}(0)z^2)G=0
\end{equation}
which has solutions of the form
\begin{equation}\label{eq:Gequation}
G(z)=a(z+b),\quad\mathrm{where}\quad b=\begin{cases}
-\frac{3(1+Q^2)}{4Q^2} & \mathrm{for}\ \Phi_+,\\
\qquad 0 & \mathrm{for}\ \Phi_-.
\end{cases}
\end{equation}
Here, $a$ is a constant we can specify by a choice of $FG$ at the horizon.  The resulting equation for $F$ is
\begin{gather}
F'' + \left(\frac{2\alpha}{z-1}+\frac{f'}{f}+\frac{2G'}{G}\right)F' + \left[\frac{\alpha}{z-1}\left(\frac{\alpha-1}{z-1}+\frac{f'}{f}+\frac{2G'}{G}\right)\right.\nonumber\\
\left.+\frac{Q^2}{f^2}(\wn^2-\qn^2f)-2Q^2\tilde{g}_{\pm}(\qn)\frac{z}{f}\right]F=0\label{eq:Fequation},
\end{gather}
where we have defined
\begin{equation}
\alpha(\wn)=\frac{-i\wn \mu}{4\pi T}\quad\mathrm{and}\quad\tilde{g}_{\pm}(\qn) = g_{\pm}(\qn) - g_{\pm}(0).
\end{equation}

Now that we have defined $G$ in this way, we can write $F$ as a series in $\wn$ and $\qn^2$, i.e.\
\begin{equation}\label{eq:Fseries}
F(z) = F_0 + \wn F_1(z) + \qn^2 F_2(z) + O(\wn^2,\wn\qn^2),
\end{equation}
then collect terms in (\ref{eq:Fequation}) at each order.  Before moving on to find $F_1$ and $F_2$, we note two points which guide the analysis.  First, the infalling boundary condition at $z=1$ implies that $F$ be regular there, and second, we only require the small $z$ expansion of $F$ in order to read off $\hat{\Pi}_{\pm}$ (as described in Appendix~\ref{app:Numerical}).

Collecting terms at $O(\wn)$, the equation for $F_1$ is
\begin{equation}\label{eq:F1equation}
F_1'' + \left(\frac{f'}{f}+\frac{2G'}{G}\right)F_1' + \frac{\alpha F_0}{\wn(z-1)}\left(-\frac{1}{z-1}+\frac{f'}{f}+\frac{2G'}{G}\right)=0,
\end{equation}
which can be integrated once to give
\begin{equation}
F_1'= -\frac{\alpha F_0}{\wn}\left(\frac{1}{z-1}+\frac{c_1 a^2}{f G^2}\right).
\end{equation}
The constant $c_1$ is determined by demanding that the $O(z-1)^{-1}$ terms cancel, and is found to be
\begin{equation}
c_1=\begin{cases}
-\frac{(3-Q^2)^3}{16Q^4} & \mathrm{for}\ \Phi_+,\\
(3-Q^2) & \mathrm{for}\ \Phi_-.
\end{cases}
\end{equation}
The factor of $a^2$ has been extracted because $F$ cannot depend on $a$.

Collecting terms at $O(\qn^2)$, the equation for $F_2$ is
\begin{equation}\label{eq:F2equation}
F_2'' + \left(\frac{f'}{f}+\frac{2G'}{G}\right)F_2' - \frac{Q^2F_0}{f}(1+2\tilde{g}_{\pm}(1))=0,
\end{equation}
which can be integrated once to give
\begin{equation}
F_2'= \frac{Q^2F_0}{f (z+b)^2}\left[c_2 + b\left(bz+z^2+2\tilde{g}_{\pm}(1)\left(\frac{bz^2}{2}+\frac{2z^3}{3}\right)\right)+\frac{z^3}{3}+\frac{\tilde{g}_{\pm}(1)z^4}{2}\right].
\end{equation}
The constant $c_2$ is determined by demanding that the $O(z-1)^{-1}$ terms cancel, and is found to be
\begin{equation}
c_2=\begin{cases}
-\frac{27+63Q^2+29Q^4+9 Q^6}{48 Q^4 \left(1+Q^2\right)} & \mathrm{for}\ \Phi_+,\\
\qquad -\frac{1}{3(1+Q^2)} & \mathrm{for}\ \Phi_-.
\end{cases}
\end{equation}

To extract $\hat{\Pi}_{\pm}$, we need only keep terms up to $O(z)$ in ${\Phi}_{\pm}$.  Let's focus on $\hat{\Pi}_+$ for now.  Expanding each factor to this order we obtain
\begin{equation}
\begin{aligned}
\Phi_+ &= (-1)^{\alpha} (1-\alpha z) a \left[F_0 + \wn (F_1(0)+F_1'(0)z) + \qn^2 (F_2(0)+F_2'(0)z) \right] (z+b) \\
&\phantom{=\ }+ O(z^2,\wn^2,\wn\qn^2).
\end{aligned}
\end{equation}
By comparison with (\ref{eq:falloff}) and substituting for $b$ we obtain
\begin{equation}
\hat{\Pi}_+ = -\frac{4 Q^2}{3 (1+Q^2)} + i\wn \left(\frac{Q}{3-Q^2} - \frac{iF_1'(0)}{F_0}\right) + \qn^2 \frac{F_2'(0)}{F_0} + O(\wn^2,\wn\qn^2).
\end{equation}
Substituting for $F_1'(0)$ and $F_2'(0)$ we obtain the final result
\begin{equation}\label{eq:Piplushydro}
\hat{\Pi}_+ = -\frac{4 Q^2}{3 (1+Q^2)} + i\wn \frac{Q\left(3-Q^2\right)^2}{9 \left(1+Q^2\right)^2} - \qn^2 \frac{Q^2\left(27+63Q^2+29Q^4+9 Q^6\right)}{27 \left(1+Q^2\right)^3} + O(\wn^2,\wn\qn^2).
\end{equation}

The case of $\hat{\Pi}_-$ is slightly different because the $z$-dependence of $F_1$ and $F_2$ is different.  We find
\begin{align}
F_1 & =-\frac{\alpha F_0}{\wn}\left(\frac{Q^2-3}{z}+d_1\right)+O(z),\\
F_2 & =Q^2F_0\left(\frac{1}{3(1+Q^2)z}+d_2\right)+O(z),
\end{align}
where $d_1$ and $d_2$ are constants.  Using $b=0$ we obtain
\begin{equation}
\Phi_- = (-1)^{\alpha}  a F_0 \left[ \alpha(3-Q^2) + \frac{\qn^2 Q^2}{3(1+Q^2)} + (1-\alpha d_1+\qn^2 Q^2 d_2)z \right] + O(z^2,\wn^2,\wn\qn^2)
\end{equation}
and thus
\begin{equation}\label{eq:Piminushydro}
\hat{\Pi}_- = \frac{1-\alpha d_1+\qn^2 Q^2 d_2}{\alpha(3-Q^2) + \frac{\qn^2 Q^2}{3(1+Q^2)}}+ O(\wn^2,\wn\qn^2).
\end{equation}

We now provide a check on our hydrodynamic result (\ref{eq:Piminushydro}).  In Section~\ref{ssec:Pole} we identified the lowest pole of $\hat{\Pi}_-$ as belonging to the hydrodynamic mode.  The location of the pole gives the dispersion relation for this mode.  Fluctuations transverse to the direction of momentum flow, as in the shear channel we are studying, excite diffusive modes.  As such, these have a dispersion relation of the form
\begin{equation}\label{eq:dispersion}
\wn = - i \mu D \qn^2
\end{equation}
with diffusion constant $D$.  This is indeed what we find in (\ref{eq:Piminushydro}) if we identify
\begin{equation}\label{eq:DiffusionConstant}
D = \frac{L^2}{3(1+Q^2)r_0}.
\end{equation}
Now, in a hydrodynamic system which is also conformal, we expect the following relations to hold for energy-momentum transport:
\begin{equation}
D = \frac{\eta}{\epsilon+P},\quad \mathrm{with}\quad P=\frac{\epsilon}{2},
\end{equation}
where $\eta, \epsilon, P$ are respectively the shear viscosity, energy density and pressure of the system.  Using the thermodynamics of our black hole (as quoted in \cite{Edalati:2010hk}),
\begin{equation}
\epsilon = \frac{r_0^3}{\kappa_4^2 L^4} (1+Q^2),\quad s=\frac{2\pi}{\kappa_4^2}\left(\frac{r_0}{L}\right)^2,
\end{equation}
we obtain
\begin{equation}
\frac{\eta}{s}=\frac{1}{4\pi}.
\end{equation}
This result is expected on general grounds for a theory dual to a two-derivative gravity theory \cite{Buchel:2003tz}.


Note that this $D$ can only be interpreted as the diffusion constant in the regime where hydrodynamics gives an effective description of the dynamics, i.e.\ for $T\gg \mu$.  Outside this regime, we expect a different effective theory to be valid.  In particular, we should not trust (\ref{eq:DiffusionConstant}) for $T \ll \mu$: either the dispersion relation needs modifying, or the other poles in $\hat{\Pi}_-$ cannot be ignored, or both.

\begin{figure}[htb]
\begin{center}
\includegraphics[width=0.5\textwidth]{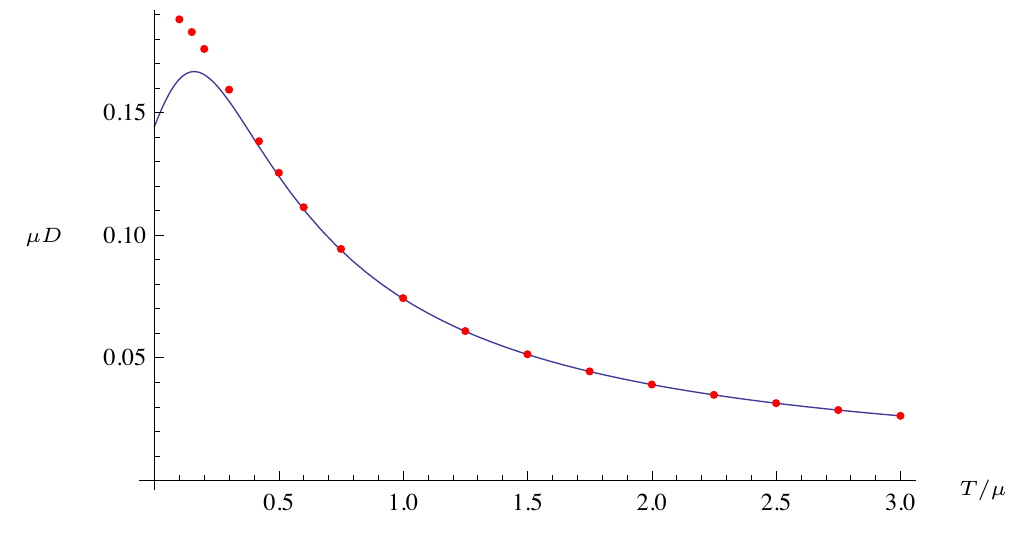}
\caption{\it Plot of $\mu D$ as a function of $T/\mu$.  The solid curve is the analytical result from \normalfont(\ref{eq:DiffusionConstant}) \it and the points were extracted from the motion of the lowest quasinormal frequency.}
\label{fig:Diffusion}
\end{center}
\vskip -2em
\end{figure}

We can calculate the diffusion constant numerically by fitting the location of the lowest quasinormal frequency of $\hat{\Pi}_-$ to the form (\ref{eq:dispersion}).   Our numerical results agree with the analytical result in (\ref{eq:DiffusionConstant}) for $T\gg \mu$, as shown in Figure~\ref{fig:Diffusion}.  To reiterate: to calculate $D$ we only need knowledge of the \emph{lowest} quasinormal frequency and not our more complete numerical results for the correlators.

\begin{figure}[htb]
\begin{center}
\includegraphics[width=0.95\textwidth]{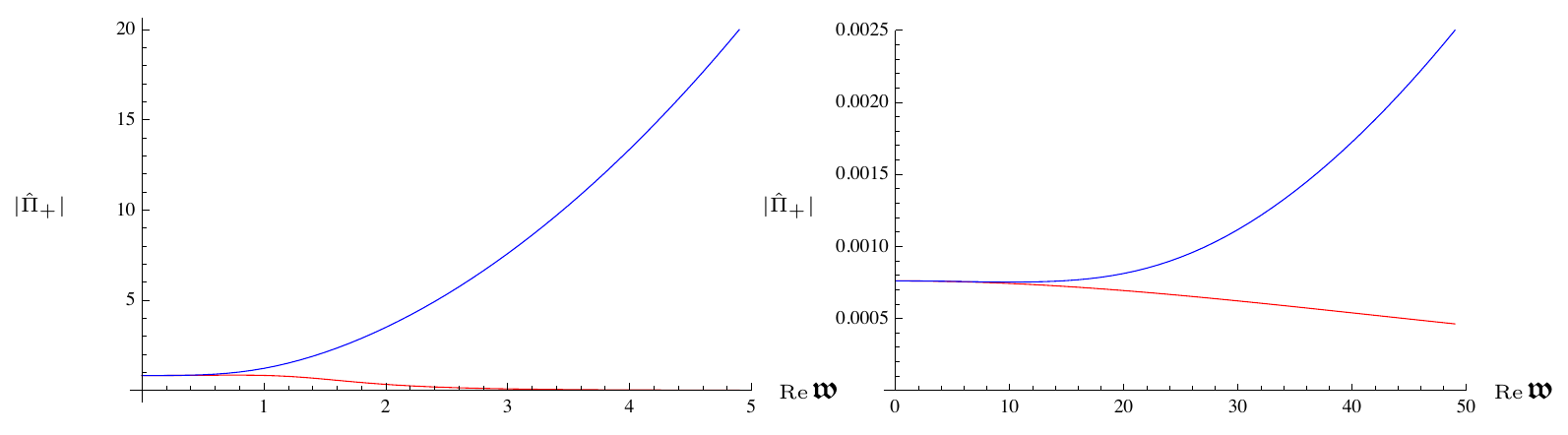}
\caption{\it A comparison between numerical (red) and analytical (blue) results for $|\hat{\Pi}_+|$.  The left plot is at $T/\mu=0.09$, the right plot is at $T/\mu=10$, and both have $\qn=10^{-6}$.  Note that we have subtracted off the large $\wn$ behaviour from all results, as discussed in Section~\ref{ssec:Behaviour}.}
\label{fig:Departure}
\end{center}
\vskip -2em
\end{figure}

Now we show how this agreement breaks down beyond the hydrodynamic regime, and as such our results are essential.  In Figure~\ref{fig:Departure} we get the correct contact term for $\hat{\Pi}_+$ as in (\ref{eq:Piplushydro}) and the results are in good agreement over a small range in $\wn$, but they soon deviate.  The agreement persists over a larger range of $\wn$ if we work at a larger value of $T/\mu$.\footnote{We used numerical integration to obtain the $O(\wn^2)$ correction to \normalfont(\ref{eq:Piplushydro}).}



\providecommand{\href}[2]{#2}\begingroup\raggedright\endgroup

\end{document}